\documentclass[conference]{IEEEtran}
\IEEEoverridecommandlockouts
\usepackage{todonotes}
\usepackage{cite}
\usepackage{amsmath,amssymb,amsfonts}
\usepackage{algorithmic}
\usepackage{graphicx}
\usepackage{textcomp}
\usepackage{xcolor}
\usepackage{multirow}
\def\BibTeX{{\rm B\kern-.05em{\sc i\kern-.025em b}\kern-.08em
    T\kern-.1667em\lower.7ex\hbox{E}\kern-.125emX}}

\usepackage{listings}
\usepackage{color}
\definecolor{codegreen}{rgb}{0,0.6,0}
\definecolor{codegray}{rgb}{0.5,0.5,0.5}
\definecolor{codepurple}{rgb}{0.58,0,0.82}
\definecolor{backcolour}{rgb}{0.95,0.95,0.92}
 
\lstdefinestyle{mystyle}{
    backgroundcolor=\color{backcolour},   
    commentstyle=\color{codegreen},
    keywordstyle=\color{magenta},
    numberstyle=\tiny\color{codegray},
    stringstyle=\color{codepurple},
    basicstyle=\footnotesize,
    breakatwhitespace=false,         
    breaklines=true,                 
    captionpos=b,                    
    keepspaces=true,                 
    numbers=left,                    
    numbersep=5pt,                  
    showspaces=false,                
    showstringspaces=false,
    showtabs=false,                  
    tabsize=2
}
 
\begin{document}
\title{A Study of Single and Multi-device Synchronization Methods in Nvidia GPUs}

\author{
  \IEEEauthorblockN{
    Lingqi Zhang\IEEEauthorrefmark{1},
    Mohamed Wahib\IEEEauthorrefmark{2}\IEEEauthorrefmark{4},
    Haoyu Zhang\IEEEauthorrefmark{3},
    Satoshi Matsuoka\IEEEauthorrefmark{4}\IEEEauthorrefmark{1},
  }
  \IEEEauthorblockA{
    \small{
      \IEEEauthorrefmark{1}
      Tokyo Institute of Technology,
      \texttt{zhang.l.ai@m.titech.ac.jp}
    }
  }
  \IEEEauthorblockA{
    \small{
      \IEEEauthorrefmark{2}
      National Institute of Advanced Industrial Science and Technology,
      \texttt{mohamed.attia@aist.go.jp}
    }
  }
  \IEEEauthorblockA{
    \small{
      \IEEEauthorrefmark{3}
      miHoYo Inc, (This work was done while the co-author worked in Tokyo Institute of Technology)
      \texttt{lynkzhang@gmail.com}
    }
  }
  \IEEEauthorblockA{
    \small{
      \IEEEauthorrefmark{4}
      RIKEN Center for Computational Science,
      \texttt{matsu@acm.org}
    }
  }
}
\maketitle

\begin{abstract}
GPUs are playing an increasingly important role in general-purpose computing. Many algorithms require synchronizations at different levels of granularity in a single GPU. Additionally, the emergence of dense GPU nodes also calls for multi-GPU synchronization. Nvidia's latest CUDA provides a variety of synchronization methods. Until now, there is no full understanding of the characteristics of those synchronization methods. This work explores important undocumented features and provides an in-depth analysis of the performance considerations and pitfalls of the state-of-art synchronization methods for Nvidia GPUs.
The provided analysis would be useful when making design choices for applications, libraries, and frameworks running on single and/or multi-GPU environments. We provide a case study of the commonly used reduction operator to illustrate how the knowledge gained in our analysis can be useful. We also describe our micro-benchmarks and measurement methods.
\end{abstract}

\begin{IEEEkeywords}
CUDA Barrier, Synchronization, GPUs
\end{IEEEkeywords}

\section{Introduction}

GPUs have been playing an increasingly important role in general-purpose computing. Different scientific areas exploit the power of GPUs to accelerate science and engineering applications. Many complex algorithms require different levels of synchronizations, through the use of barriers. Until recently~\footnote{Nvidia introduced a hierarchy of synchronization methods (based on Cooperative Groups(CG)) since CUDA 9.0~\cite{nvidia2019programming}}, developers used two methods of synchronization in CUDA. First, developers made use of CUDA thread block synchronization to develop complex algorithms~\cite{harris2007optimizing}. Second, for applications like Deep Learning (DL), the CPU-side implicit barrier occurring after the kernel launch function is used for device-wide synchronization~\cite{tokui2015chainer}. 

Due to the importance of device-wide synchronization, several researchers attempted to develop software device-wide barriers~\cite{sorensen2016portable,xiao2010inter}. Liu et al.~\cite{liu2018efficient} also proposed a hardware-software cooperative framework for synchronization. Yet the increase in complexity and density of GPUs in GPU-based systems, e.g. Nvidia DGX-2 includes $16$ GPUs, call for a general and high-performance method for devices-wide and multi-GPU synchronization. Recently Nvidia proposed methods for synchronizations that spans all levels of granularity from a small group of threads in a GPU to a multi-GPU device: warp level, thread block level, and grid level. The grid level synchronization can be a productive way to perform device-wide and multi-device level synchronization. This hierarchy of synchronization methods can make GPU programming more productive. Thus, it is important to study the performance characteristics of different levels of synchronization methods. 

In this paper, we characterize the synchronization methods in Nvidia GPUs. Specifically, in this work:

\begin{itemize}
    \item We identify the performance characteristics of different synchronization methods in Nvidia GPUs.
    \item We use different implementations of the reduction operator as a motivating example to demonstrate how to use the knowledge gained in this study to optimize the reduction kernel. 
    \item We explore the pitfalls of using several synchronization instructions. 
    \item We provide our micro-benchmarks used in measurements~\footnote{The source code is available at: \\https://github.com/neozhang307/SyncMicrobenchmark}.
\end{itemize}

\section{Background}

\subsection{CUDA Programming Model}


CUDA is a C-like programming model for Nvidia GPUs. It offers three levels of programming abstractions: thread, thread block, and grid. Among them, thread is the most basic programming abstraction. At the hardware side, there is a hierarchy that maps to the CUDA programming model. Three different levels of hardware resources exist: ALU, Stream Multi-Processor (SM), and the GPU. Take the Volta V100 \cite{nvidia2017v100} as an example, a V100 GPU consists of $80$ SMs; an SM is partitioned into $4$ processing blocks, each consists of several ALUs, e.g. $16$ FP32 Cores. 

A \emph{warp} in CUDA is a small number of threads executed together as a working unit in a SIMT fashion. A warp in all Nvidia GPU generations consists of $32$ threads. Inside an SM in V100 there are $4$ warp schedulers corresponding to the $4$ partitions inside one SM. CUDA's runtime will schedule one thread block to only one SM, and one grid to only one GPU, though it may occupy several SMs. 

\begin{figure}[t]
\centering
\includegraphics[width=\linewidth]{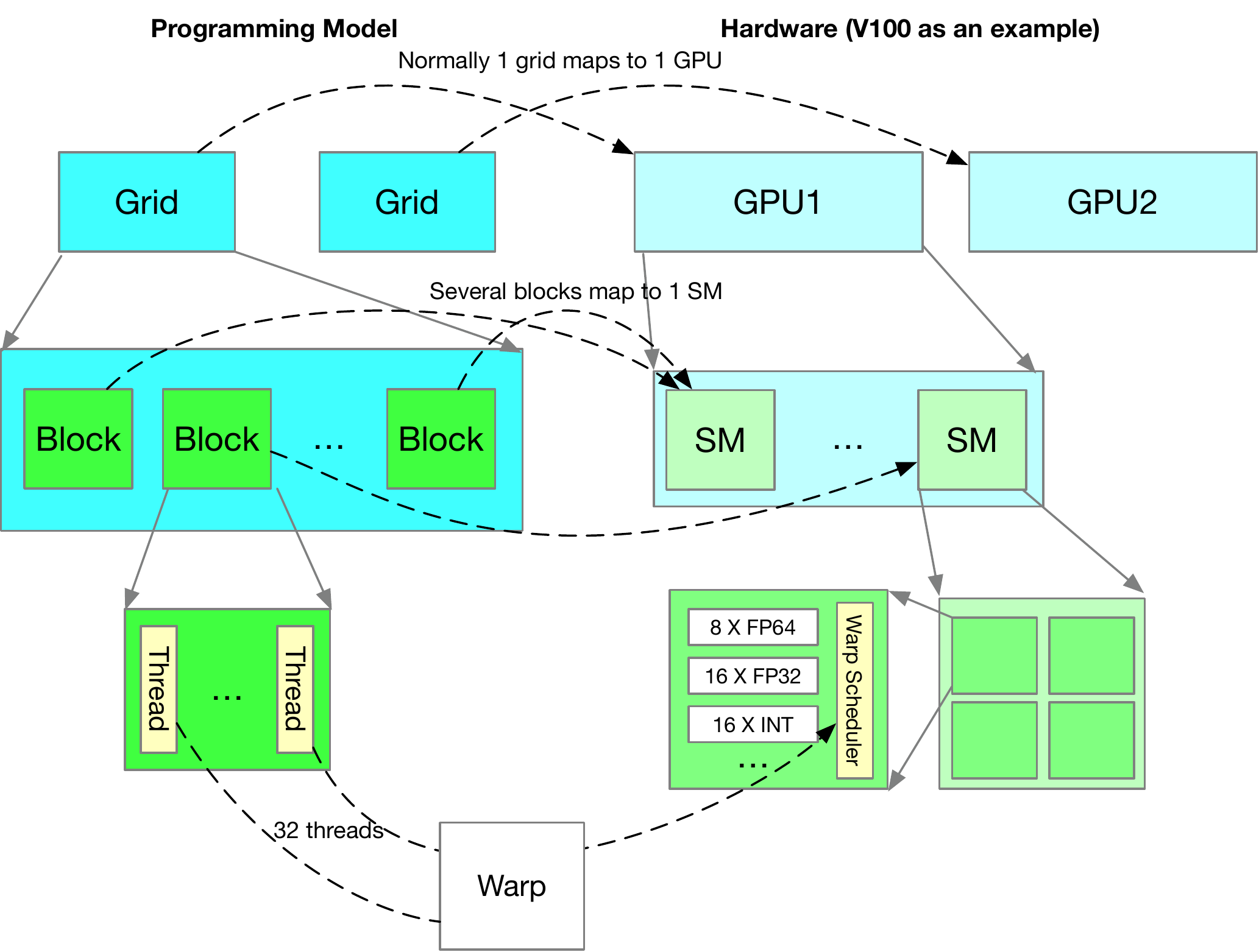}
\caption{CUDA programming model and corresponding hardware structure}
\label{Fig:programming_model}
\end{figure}

Figure \ref{Fig:programming_model} shows the details of CUDA programming model, its corresponding hardware abstraction, and the mapping relationship between them.

\subsection{Related Work}
Many efforts have been done to micro-benchmark GPUs. Volkov et al.~\cite{volkov2008benchmarking} benchmarks were partially used to study kernel launch overhead, manual device-wide barriers, data transfer, pipeline latency, instructions throughput, and metrics related to GPU memory system. This knowledge discovered was then used to tune several dense linear algebra algorithms. Wong et al.~\cite{wong2010demystifying} proposed the use of more fine-grained micro-benchmarks to understand the performance of GPUs, including the behavior of instructions and memory structure of GPUs. Zhang et al.~\cite{zhang2017understanding} introduced assembly-level micro-benchmarks. Recently, Jia et al. use ASM code to run micro-benchmarks on new Nvidia Tesla GPUs, i.e. V100 and P100~\cite{jia2018dissecting}. Several other works mainly focused on the memory hierarchy of GPUs, e.g.~\cite{baghsorkhi2012efficient,mei2014benchmarking,mei2016dissecting}. Among them, Mei et al. ~\cite{mei2016dissecting} discovered some cache patterns that were missed by previous researches. To the authors' knowledge, none of the GPU micro-benchmarking efforts focus on CUDA's hierarchy of synchronizations. 


Volkov et al.~\cite{volkov2010better} also compared kernel launch overhead and a manually implemented software barrier. Yet they only tested the overhead of light kernels, which is not practical for most of the applications. Other efforts analyzed software synchronization methods by comparing the performance of implementations of several algorithms with and without their software synchronization methods~\cite{sorensen2016portable,xiao2010inter,li2015fine}. The analysis works on case-by-case bases and can not be generalized to different kernels. 

\section{Overview of Synchronization Methods in Nvidia GPUs}
\subsection{Primitive Synchronization Methods in Nvidia GPUs}
Starting from CUDA 9.0, Nvidia added the feature of \emph{Cooperative Groups (CG)}. This feature is planned to allow scalable cooperation among groups of threads and provide flexible parallel decomposition. Coalesced groups and tile groups can be used as a method to decompose thread blocks. Beyond the level of thread blocks, grid synchronization is proposed for inter-block synchronization. Multi-grid synchronization is proposed for inter-GPU synchronization.

In the current version of CUDA (10.0), tile group and coalesced group only work correctly inside a warp. Analysis of PTX code shows that those two instructions are transformed to the \emph{warp.sync} instruction. Hence, as it stands, we consider the synchronization capability of those methods to be only applicable to the warp level. 

\begin{figure}[t]
\centering
\includegraphics[width=\linewidth]{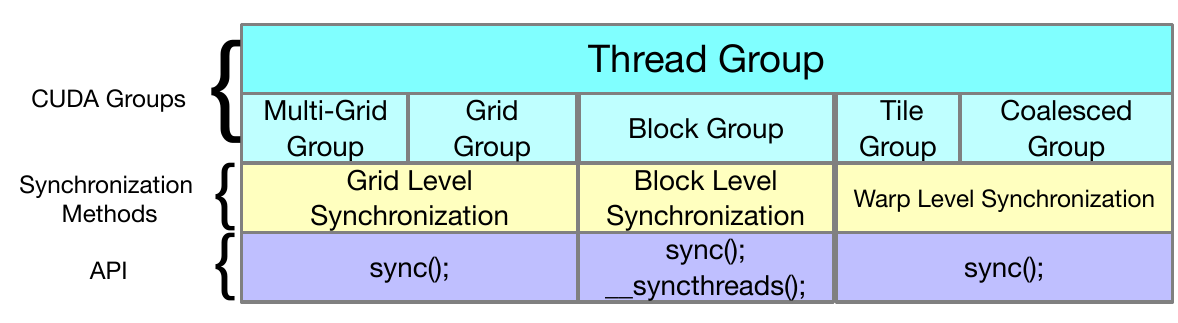}
\caption{Hierarchy of synchronizations in CUDA}
\label{Fig:sync_structure}
\end{figure}

Figure \ref{Fig:sync_structure} shows the granularity of cooperative groups and synchronization in the current version of CUDA.
\subsubsection{Warp-level Synchronization (Synchronization Inside a Single GPU)}
Current CUDA supports two intra-warp synchronization methods, i.e. tile synchronization and the coalesced group synchronization corresponding respectively to the tile group and coalesced group in Figure \ref{Fig:sync_structure}. Previous versions of CUDA guarantee that all threads inside a warp process the same instruction at a time. Yet the introduction of synchronization methods inside a warp plus the fact that each thread now has its own Program Counter (PC) implies a future possibility of removing this feature.
\subsubsection{Block-level Synchronization (Synchronization Inside a Single GPU)} 
Block-level synchronization corresponds to the thread block in the programming model. According to CUDA's programming guide~\cite{nvidia2019programming}, its function is the same as the classical synchronization primitive \_\_syncthreads(). 

\subsubsection{Grid-level Synchronization (Single GPU Synchronization)}
Starting from CUDA 9.0, Nvidia introduced grid group grid-level synchronization. Grid-level synchronization is a method to do single GPU synchronization. In order to use a grid group, cudaLaunchCooperativeKernel() API call is necessary, in comparison to the traditional kernel launch ($<<<>>>$). 
\subsubsection{Multi-Grid Level Synchronization (Multi-GPU Synchronization)} CUDA 9.0 also introduced the concept of multi-grid group. This group is initialized by a kernel launch API: cudaLaunchCooperativeKernelMultiDevice(). Synchronizing this group can act as a way to do multi-GPU synchronization in a single node. 
\subsection{Non-primitive Synchronization}
\subsubsection{Software Barrier for Synchronization}
Li etc.~\cite{li2015fine} researched fine-grained synchronization. Beyond it, Xiao, etc.~\cite{xiao2010inter} introduced a software device-level synchronization. The authors limit the number of blocks per SM to only one in order to avoid deadlocks. Sorensen et al. extended this work by adding an automatic occupancy discovery protocol to discover activate warps~\cite{sorensen2016portable}.
\subsubsection{Implicit Barrier for Synchronization}
Before the introduction of grid-level synchronization, the typical way to introduce a device-wide barrier to a program was to use several kernels in a single CUDA stream. A stream is a logical queue that enforces an execution order on the CUDA kernels in the stream, i.e. the kernels and data movement commands are executed in the order by which they appeared in the stream. For example, many DL frameworks, e.g., Chainer~\cite{tokui2015chainer}, use this method to enforce execution order.
\subsubsection{Multi-GPU Synchronization}
The common way to do multi-GPU synchronization is to synchronize CPU threads orchestrating the GPUs. The basic idea is to use one CPU thread per device (or one MPI rank per device). Additionally, with the help of the \emph{GPUDirect} CUDA technology, it is also possible to implement multi-GPU software barriers using GPUDirect APIs. 

Since we are concerned in this paper with studying general and intrinsic barrier methods, we would not discuss manually implementation barriers, including software barriers and GPUDirect based manually implementations. 

\section{Synchronization via CPU-side Implicit Barriers}
\label{sec:implicit}
Launching new kernels in a single stream can act as a device-wide implicit barrier to maintain the order of the program. Yet launching an additional kernel is not a free lunch: it will also introduce overheads. This section will inspect the overhead of traditional launch function, i.e., the $<<<>>>$ kernel invocation method, and the new launch functions, i.e. cudaLaunchCooperativeKernel() and  cudaLaunchCooperativeKernelMultiDevice() Nvidia introduced from CUDA 9.0 for CG. 

To simplify our discussion, this section does not consider the extra overhead of launching the first kernel. Instead, in all our measurements we assume a warm-up kernel was already launched, and we focus our analysis on the behavior of kernels launched after the warm-up kernel.

Before further discussion in this section, we introduce the following terms:
\begin{itemize}
\item \textbf{Kernel Execution Latency:} Total time spent in executing the kernel, excluding any overhead for launching the kernel.
\item \textbf{Launch Overhead:} Latency that is not related to kernel execution. 
\item \textbf{Kernel Total Latency:} Total latency to run kernels. $T_{Kernel\ Total\ Latency}=T_{Kernel\ Execution\ Latency}+T_{Launch\ Overhead}$
\end{itemize}

\begin{figure}[t]
\centering
\begin{lstlisting}[language=C,gobble=2]
    __global__ void null_kernel(){
        //kernel execution latency is 10 us here. 
        repeat10(asm volatile("nanosleep.u32 1000;");)
    }
    ...
    record(timer1);
    repeat1(launch(null_kernel, launchparameters););
    cudaDeviceSynchronize();
    record(timer2);
    repeat5(launch(null_kernel, launchparameters););
    cudaDeviceSynchronize();
    record(timer3);
    ...
\end{lstlisting}
\caption{Sample code to micro-benchmark implicit barriers for a null (empty) kernel}
\label{Fig:code_example}
\end{figure}

Figure \ref{Fig:code_example} is our sample code for micro-benchmarks. It also shows the concept of kernel execution latency and kernel total latency. Kernel execution latency is controlled by the sleep instruction. $T_{kerne\ totall\ latency}=((timer3-timer2)-(timer2-timer1))/(5-1)$ here; Elaborate details on the bench-marking methods are discussed in Section~\ref{sec:benchimplicit}.  

In this way, We measured the launch overhead by using the kernel fusion method. We also test the kernel total latency of a null kernel for comparison. Table \ref{tab:lgovh} shows the result. 

\begin{table}[t]
\centering
  \caption{Launch Overhead and Null kernel Latency of Different Launch Functions}
  \label{tab:lgovh}
  \begin{tabular}{l|l|l}\hline
   \hfil &&\hfil  Null Kernel\\
      \hfil Launch Type&Launch Overhead& Kernel Total Latency\\
   \hfil &\hfil  (ns)&\hfil (ns)\\\hline
   \textbf{ Traditional} &\hfil 1081&\hfil  8888\\
   \textbf{ Cooperative} &\hfil 1063&\hfil 10248\\
    \textbf{Cooperative Multi-Device} &\hfil 1258&\hfil 10874\\\hline
\end{tabular}
\end{table}

\section{Single GPU Synchronization}
\label{sec:sing}
In this section, we characterize the performance of warp, thread block, and grid level synchronization. Warp and block abstractions exist inside an SM. For warp and block, we used the micro-benchmark discussed in Section~\ref{sec:benchintrasm}. Grid is an inter-SM abstraction, for that, we used the micro-benchmark discussed in Section~\ref{sec:benchintersm}. 

For the warp shuffle operation and block synchronization operation, the throughput is reported by CUDA programming guide~\cite{nvidia2019programming} at the granularity of warps and blocks, respectively. Yet it is possible that the size of a group that performs synchronization or shuffle would influence the performance itself. Hence in this work, we consider the group size when experimenting with warp shuffle and block synchronization. 

\subsection{Warp-Level Synchronization}
\label{sec:warprst}
The current CUDA (10.0) supports two kinds of warp level synchronization: tile group based and coalesced group based (as seen in Figure \ref{Fig:sync_structure}). Additionally, the CUDA shuffle operation, which exchanges a register value among threads in a warp, is an operation that implies a synchronization after it. We also include the results of the shuffle operation. 

Since the size of a synchronization group might influence the result, we tested every possible group size for both tile group and coalesced group. The possible tile group sizes are: $1$, $2$, $4$, $8$, $16$, and $32$. The possible coalesced group size ranges from $1$ to $32$. Latency is tested by using only $32$ threads (a warp) in a CUDA kernel with one block. The throughput is tested by iterating every possibility pair of up to $1024$ threads and up to $64$ blocks per SM and recording only the highest result. Table \ref{tbl:warprst} shows the result of warp level synchronization.

For tile group synchronization the size of the group influence neither latency nor throughput. A possible explanation is that CUDA could be merging all the concurrent tile group synchronization instructions into a single instruction. For coalesced group synchronization, the group size does not influence the performance of P100. The group size does, however, influence the performance of coalesced group in V100. The performance is the highest when all the threads inside a warp belong to a single coalesced group. For convenience, because the group size doesn't influence the total latency of tile group synchronization, we only record the throughput in the case of a group size of $32$ in tile group synchronization.

\begin{table}[t]
\caption{\label{tbl:warprst}Performance of Warp Synchronization in a Block}
\centering
\begin{tabular}{l|l|l|l|l|l|l}
\textbf{Type}&\multicolumn{2}{l|}{\textbf{Latency}}&\multicolumn{2}{l|}{\textbf{Throughput}}&\multicolumn{2}{l}{\textbf{Reference}\cite{nvidia2019programming}}\\
(group size) &\multicolumn{2}{l|}{\textbf{cycle}}  &\multicolumn{2}{l|}{\textbf{(sync/cycle)}}  &\multicolumn{2}{l}{thread op/cycle}\\\hline 
  &\hfil \textbf{V100}&\hfil \textbf{P100}&\hfil \textbf{V100}    &\hfil \textbf{P100}&\hfil \textbf{V100}    &\hfil \textbf{P100}\\\hline 
\textbf{Tile}(*)            &\hfil 14   &\hfil 1    &\hfil 0.812    &\hfil 1.774&\hfil   -  &\hfil  -\\
\textbf{Shuffle(Tile)}(*)   &\hfil 22   &\hfil 31   &\hfil 0.928    &\hfil 0.642&\hfil 32    &\hfil 32\\
\textbf{Coalesced}(1-31)    &\hfil108   &\hfil1     &\hfil 0.167    &\hfil 1.791&\hfil  -   &\hfil -\\
\textbf{Coalesced}(32)      &\hfil14    &\hfil1     &\hfil 1.306    &\hfil 1.821&\hfil -    &\hfil -\\
\textbf{Shuffle(COA)}(*)    &\hfil 77   &\hfil 50   &\hfil 0.121    &\hfil 0.166&\hfil  -   &\hfil -\\\hline
\textbf{Block(warp))}       &\hfil 22   &\hfil 218      &\hfil 0.475    &\hfil 0.091&\hfil  16   &\hfil 32\\\hline
\end{tabular}
\end{table}

We use the reference throughput of shuffle operation mentioned in the CUDA programming guide~\cite{nvidia2019programming} in Table~\ref{tbl:warprst}. Apparently, the performance of V100 is closer to the theoretical result in the programming guide. On the other hand, there seem to be some overheads that influence the throughput of the shuffle operation in P100. 

\subsection{Block-Level Synchronization}

We tested every possible group size at the block level, i.e. starting from 32 to 1024. We find that the throughput of block-level synchronization is related to the number of active warps per SM. 

\begin{figure}[t]
\centering
\begin{minipage}{0.47\textwidth}
\centering
\includegraphics[width=0.9\textwidth]{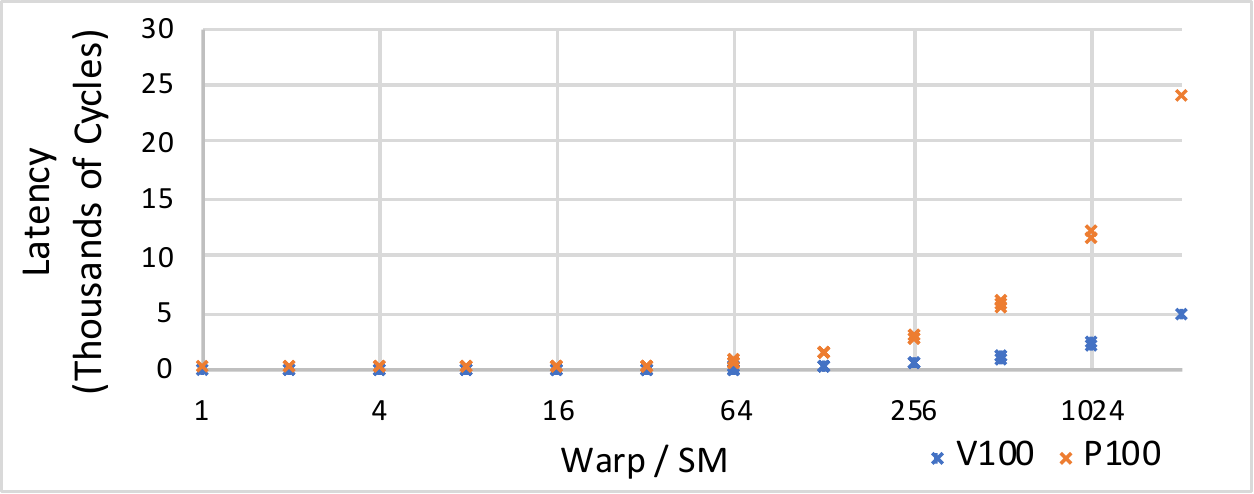}
\end{minipage}
\begin{minipage}{0.47\textwidth}
\centering
\includegraphics[width=0.9\textwidth]{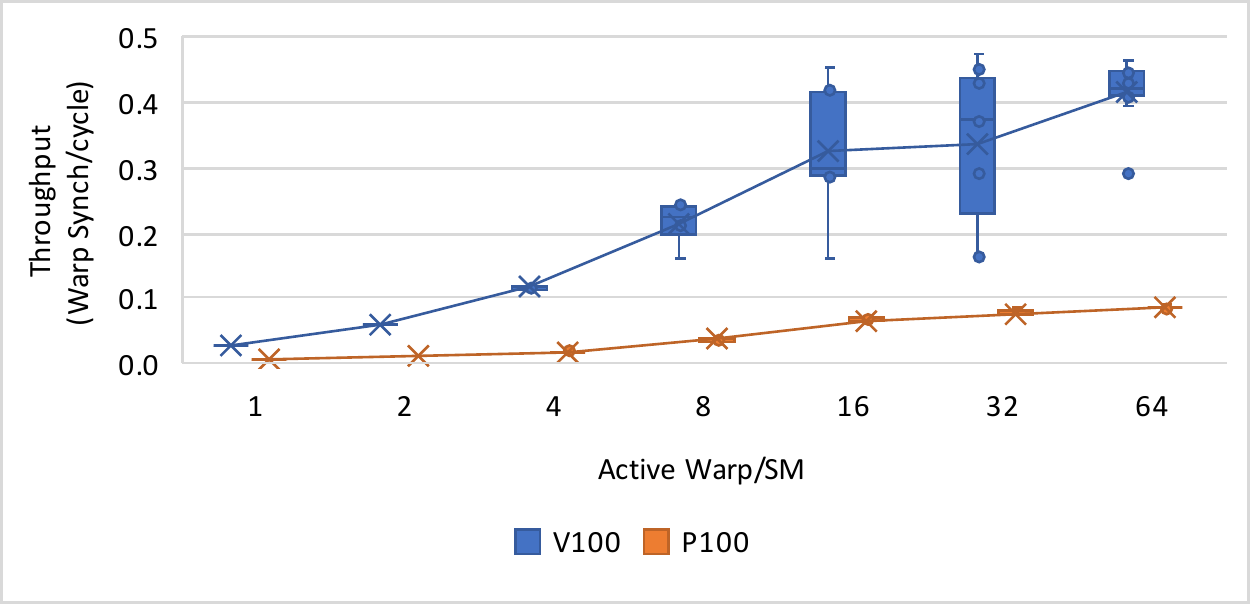}
\end{minipage}
\caption{Relationship between throughput of block sync (per warp perspective) (up) and active warp/SM perspective (down)}
\label{Fig:warppsm}
\end{figure}

Figure~\ref{Fig:warppsm} shows the relationship between the throughput of block synchronization divided by warp count (warp sync per cycle) and the maximum number of activate warps per SM (as calculated by~\cite{nvidia2017v100}). When the warp count exceeds the size of max activate warp per SM, the device is saturated and the throughput of block synchronization reaches its maximum. 

With this observation, we conclude that the performance of block-level synchronization is related to the warp count per SM. We further summarize the performance of block synchronization from a warp's perspective in Table~\ref{tbl:warprst}.

CUDA's programming guide \cite{nvidia2019programming} reports that the throughput for \_\_syncthreads() (or block-level synchronization) is $16$ operations per clock cycle for capability 7.x (V100) and $32$ for capability 6.0 (P100). The throughput of V100 is relatively close to $16$ op/cycle. But the result of P100 is far away from $32$ op/cycle. To further support our result, the inverse of the gradient of the points in the up part of Figure~\ref{Fig:warppsm} can represent throughput. Obviously, the gradient of block synchronization in P100 is larger than V100. So, the throughput of P100 should not be larger than V100's.

Admittedly, it is also possible that the performance of block synchronization in P100 is not ideal due to over-subscription. Yet the latency of block synchronization in P100 is so large that it is nearly impossible to find a point at which the instruction pipeline is saturated while the overhead of over-subscription is not so severe.

\subsection{Grid-Level Synchronization}
\begin{figure}[t]
\centering
\includegraphics[width=0.48\textwidth]{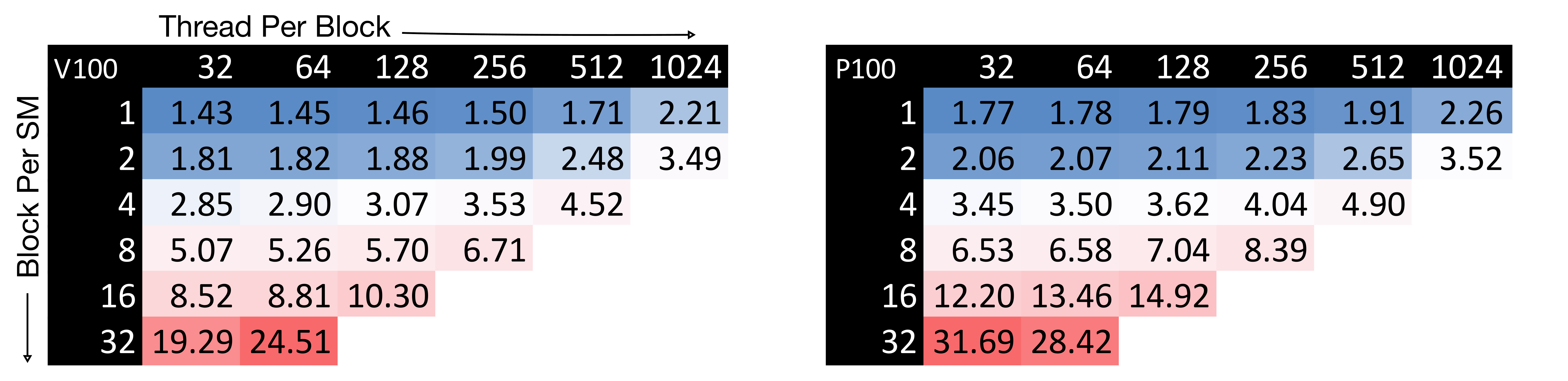}
\caption{Latency ($us$) of grid synchronization in V100 (left) and P100 (right)}
\label{Fig:gridsync}
\end{figure}

Figure \ref{Fig:gridsync} shows the heat map of grid synchronization. It shows that in both V100 and P100 the latency of grid synchronization is more related to the grid dimension (specifically, block count per SM) than to the block dimension. 

No matter how small the grid is, it seems that it is still slower than the overhead of kernel launch we measured in Section \ref{sec:implicit}. Single GPU grid synchronization might not bring about any benefit in performance, in comparison to implicit barrier methods. Yet we argue that this performance difference is negligible (at most $2.5 us$ with two blocks/SM) in real applications. In addition, using the implicit barrier instead would eliminate the possibility of data reuse in shared memory and registers.

\section{Multi-GPU Synchronization Methods}
We consider three ways to do multi-GPU synchronization:

\subsection{Using Multi-device Launch Function as an Implicit Barrier}
When using the multi-device launch function with the default flag, kernels will not execute until all the previous operations in all the GPU streams involved have finished execution~\cite{nvidia2019api}. Although this implicit barrier method is not commonly used, we nonetheless evaluate it to assess if this method is a valuable alternative. Section~ \ref{sec:benchimplicit} discusses in detail micro-benchmark we use in this subsection.

\subsection{Using CPU-side Barriers}
A common way to make a barrier between GPUs is to use CPU threads or processes to synchronize different GPUs. We use openMP to measure the overhead in this case. Each thread calls the cudaDeviceSynchronize() API to ensure the asynchronously launched GPU kernels are executed till their end. In addition, the threads use the openMP barrier API to synchronize. Figure \ref{fig:openmpcode} shows the code example for this kind of barrier. 
Finally, we appropriately pin the CPU threads. We applied the same micro-benchmark discussed in Section~\ref{sec:benchimplicit} for this subsection.

\begin{figure}[t]
\centering
\begin{lstlisting}[language=C,gobble=2]
	#pragma omp parallel num)threads(GPU_count){
		unit gid=omp_get_thread_num();
		cudaSetDevice(gid);
		...
		kernel<<<>>>();
		cudaDeviceSynchronize();
		#pragma omp barrier
		...
	}
\end{lstlisting}
\caption{\label{fig:openmpcode}Code example of using CPU threads for synchronization}
\end{figure}

\subsection{Using Multi-grid Synchronization}

\begin{figure}[t]
\centering
\includegraphics[width=0.48\textwidth]{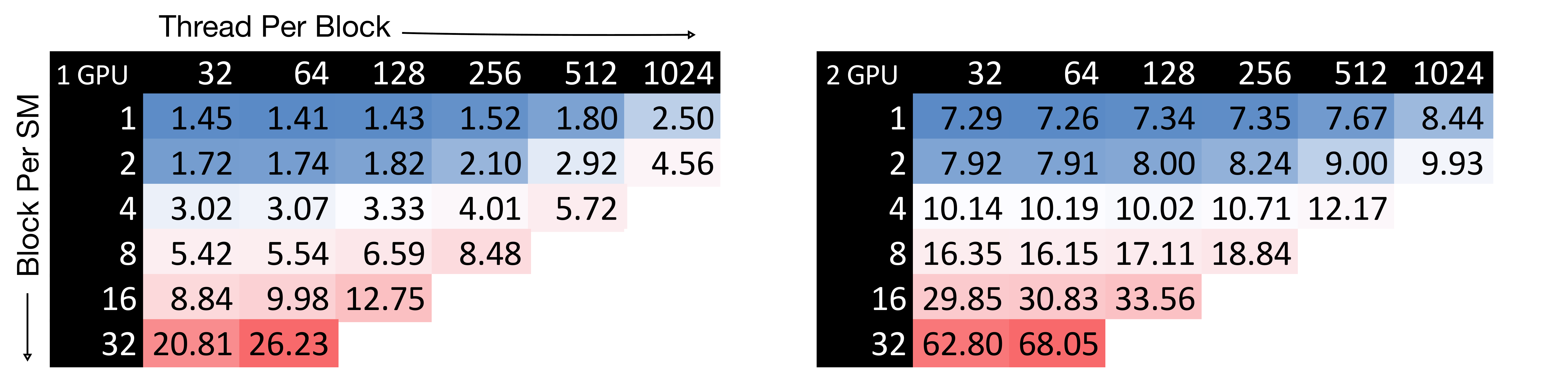}
\caption{Latency ($us$) of multi-grid synchronization in P100 platform for one GPU (left) and two GPUs (right)}
\label{Fig:mgridp100}
\end{figure}

\begin{figure}[t]
\centering
\includegraphics[width=0.48\textwidth]{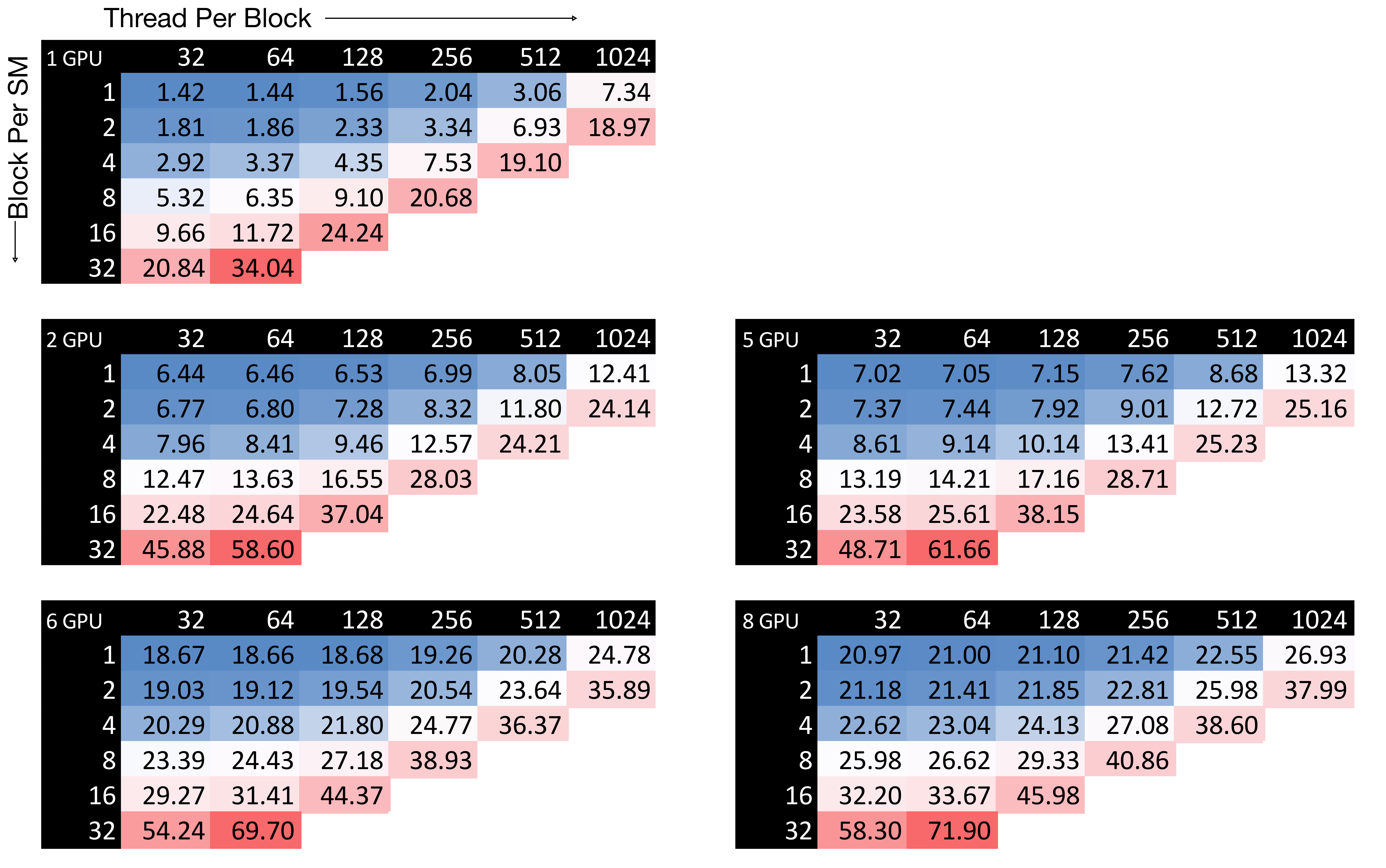}
\caption{Latency ($us$) of multi-grid synchronization in V100 platform}
\label{Fig:mgridv100}
\end{figure}

Section~\ref{sec:benchintersm} discusses in detail micro-benchmark we use in this subsection. Figure~\ref{Fig:mgridp100} and Figure~\ref{Fig:mgridv100} show the heat maps of the latency of multi-grid synchronization in V100 and P100. Because the inter-connection in the P100 system is PCIe, the performance is worse than the V100 system that is equipped with NVLink connection between devices.

We experimented with all $8$ GPUs in the DGX-1, we found that the performance of multi-grid synchronization among 2-5 GPUs is similar to each other, and the performance of multi-grid synchronization among 6-8 GPUs are similar to each other. This behaviour is likely related to the internal NVLink network structure of DGX-1. From Figures~\ref{Fig:mgridp100} and~\ref{Fig:mgridv100}, we can see that the performance of multi-grid synchronization is influenced by both the grid dimension and number of active warps per SM. With $block/SM<=8$ and $ warp/SM<=32$, the performance is acceptable. Apart from the case of one GPU, latency in all cases is no more than 2x slower than the fastest case ($1$ block/SM, $32$ threads/block) and 2x faster than the slowest case ($32$ blocks/SM, $64$ threads/block).

\subsection{Comparison}
\begin{figure}[t]
\centering
\includegraphics[width=0.48\textwidth]{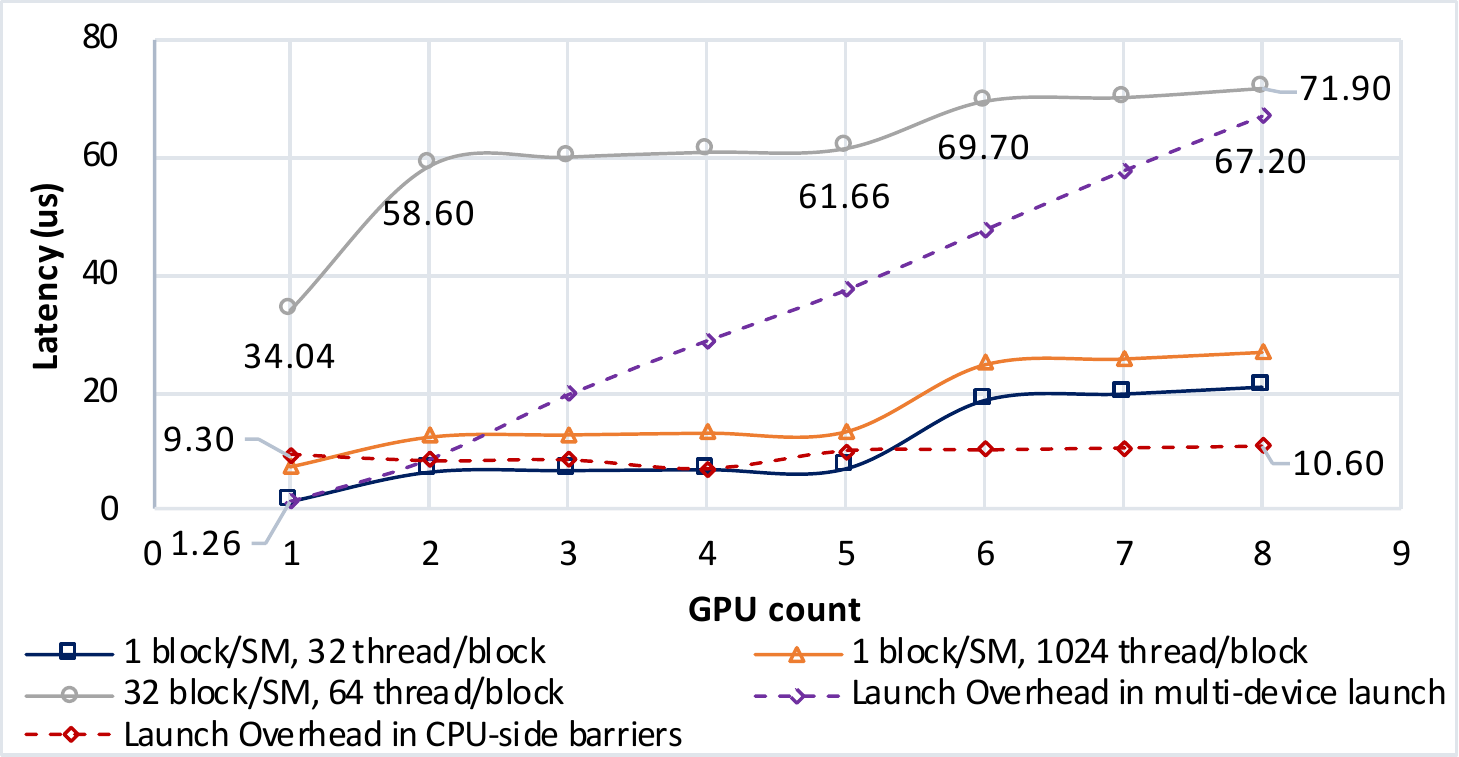}
\caption{Comparison of implicit barriers performance: multi-device launch vs. CPU-side barriers and multi-grid synchronization across 8 GPUs in DGX-1}
\label{Fig:DGX1_gpus}
\end{figure}

Figure~\ref{Fig:DGX1_gpus} shows the results of all three multi-GPU synchronization methods across $8$ GPUs in DGX-1. For simplification, we only plot the data of three cases of multi-grid synchronization in Figure~\ref{Fig:DGX1_gpus}: a) one block/SM, $32$ threads/block as the fastest case, b) $32$ blocks/SM, $64$ threads/block as the slowest case, and c) one block/SM, $1024$ threads/block as a general case, which is within the parameters we recommended in the previous paragraph.

The CPU-side barrier relying on openMP barriers outperform implicit barriers in multi-device launch when the GPU count is larger than two. Also, the overhead of the CPU-side barrier is relatively steady w.r.t. GPU count. It is worth mention that this result is relatively close to the kernel total latency of a null kernel as shown in Table~\ref{tab:lgovh}.


 Figure~\ref{Fig:DGX1_gpus} shows two performance drops in multi-grid synchronization. 
We anticipated that the second drop would be between $4$ GPUs and $5$ GPUs, based on the internal network structure of DGX-1 that groups $4$ GPU together. However, we find no reasons for the performance drop between $5$ GPU and $6$ GPU.

The figure shows that multi-grid synchronization outperforms the multi-device kernel launch function as an implicit barrier. On the other hand, as long as the program is not oversubscribed, i.e., no more than $1024$ threads per SM, the performance of multi-grid synchronization is at most 3x slower than CPU-side barriers. Yet the difference is around $16 us$, which is practically not an issue in the situation of $8$ GPUs.
We argue that this minor cost should not discourage programmers from considering the use of multi-grid synchronization in their algorithms, given the utility provided in terms of simplicity of programming, and avoiding reliance on third-party libraries such as openMP or MPI.

\section{Case Study: Reduction Operator}
We use the reduction operator (summing the elements of an array) as a case study. Harris et al.~\cite{harris2007optimizing} is a notable work that focused on optimizing the reduction operator in CUDA. They studied several optimization methods and optimized the operator by optimizing for maximum memory bandwidth utilization. Additionally, Luitjens et al.~\cite{luitjens2014faster} introduced the use of the shuffle primitive in reduction. The optimized reduction kernels can be found in CUDA SDK samples~\cite{nvidia2019sample}. There are other similar optimization strategies~\cite{martin2012algorithmic,jradi2018fast}. To the best of the authors' knowledge, all of the previous strategies didn't quantitatively compare different synchronization methods in different implementations. In this section, we will demonstrate how to capitalize on the analysis in previous sections to make a decision between different reduction implementations, depending on the input size and number of workers involved. This approach can be applied to optimize any of the previous reduction implementations and many other code generation frameworks~\cite{de2019automatic}. 


In addition to using single GPU synchronization methods in optimizing for input size, there is a programmability benefit in using multi-grid synchronization for multi-GPU systems. In dense system, such as Nvidia DGX-1 and DGX-2, the peer access feature enables one GPU to access the memory of another GPU. In this case, multi-grid synchronization provides an easy way to ensure sequential consistency. We explain this in detail in section~\ref{sec:mul_red}.  

It is important to mention another potential benefit that does not appear in the case of the reduction kernel. There is a potential of improving data reuse by the means of replacing several kernel invocations with a single persistent kernel that uses multi-grid synchronization. An example of that would be replacing kernel invocations in iterative stencil methods with a persistent kernel that includes the time loop inside the kernel. 

\subsection{Performance Model}\label{sec:pmodel}

\begin{table} [t]
\caption{\label{tbl:reduce_concurency} Projected concurrency of the two configurations in Section~\ref{sec:predict}}
\centering
\begin{tabular}{l|l|ll|ll|ll}
\multicolumn{2}{l|}{\textbf{scenery}}&\multicolumn{2}{l|}{\textbf{bandwidth}}&\multicolumn{2}{l|}{\textbf{latency}}&\multicolumn{2}{l}{\textbf{concurrency}}\\
\multicolumn{2}{l|}{\textbf{}}&\multicolumn{2}{l|}{\textbf{B/cycle}}&\multicolumn{2}{l|}{\textbf{cycle}}&\multicolumn{2}{l}{\textbf{B}}\\\hline
\multicolumn{2}{l|}{}&        V100&P100&V100&P100&V100&P100\\\hline
1          & 1 thrd.     &   \hfil       0.62      &\hfil  0.43      &\hfil 13.0 &\hfil 18.5  & \hfil8  &\hfil8\\
            &1 warp      & \hfil        19.6       &\hfil  13.8     &\hfil 13.0 &\hfil 18.5  &\hfil 256&\hfil 256\\\hline
2         & 32 thrd.    &  \hfil        19.6      &\hfil  13.8     &\hfil 13.0 &\hfil 18.5   & \hfil 256&\hfil 256\\
            &1024 thrd & \hfil         215      &\hfil  141       &\hfil 13.0 &\hfil 18.5   &\hfil 2796 &\hfil 2615\\\hline
\end{tabular}
\end{table}

We assume that the throughput is indifferent to the size of the problem (for any problem size that fully utilizes the device). We also assume that the cost of synchronization is the main cost of multi-threading. We can use Equation~\ref{eqt:predict} to know when to use fewer threads. In this equation, "basic" might refer to single thread, single warp, single block, or single GPU, and "more" corresponds to more threads, more warps, more blocks, or multi-GPU. We use Little's Law \cite{little2008little} to compute concurrency (Equation~\ref{eqt:little}). To simplify the problem, we consider $T_{basic}$ as the latency in Little's Law, and $T_{more}$ includes the overhead of synchronization as Equation~\ref{eqt:latency} shows. From this equation we can imagine three different scenarios: 

\begin{enumerate}
\item If the input size is not larger than the concurrency of "basic" threads, using fewer threads would always be more profitable. 
\item If the input size is larger than the concurrency of "basic" threads and no larger than the concurrency of "more", we can use Equation~\ref{eqt:predictsize_s} to compute the switching point. 
\item If the input size is larger than the concurrency of "more" threads. We can use Equation~\ref{eqt:predictsize_l} to know at which point we should use fewer threads. 
\end{enumerate}

\begin{equation}
    \begin{aligned}
    C= T * Thr
    \end{aligned}
    \label{eqt:little}
\end{equation}
\begin{equation}
    \begin{aligned}
    T_{basic}+\tfrac{Max(0,N-C_{basic})}{Thr_{basic}}<T_{more}+\tfrac{Max(0,N-C_{more})}{Thr_{more}}
    \end{aligned}
    \label{eqt:predict}
\end{equation}
\begin{equation}
    \begin{aligned}
    T_{more}=T_{basic}+T_{sync}=T+T_{sync}
    \end{aligned}
    \label{eqt:latency}
\end{equation}
\begin{equation}
    \begin{aligned}
    N_m < (T+T_{sync})*Thr_{basic}
    \end{aligned}
    \label{eqt:predictsize_s}
\end{equation}
\begin{equation}
    \begin{aligned}
    N_l < \tfrac{(T_{sync})*Thr_{more}*Thr_{basic}}{Thr_{more}-Thr_{basic}}
    \end{aligned}
    \label{eqt:predictsize_l}
\end{equation}
$_{*\mathit{(T\ represent\ Latency; Thr\ represent\ Throughput;} }$\ \\
$_{\mathit{C\ represent\ concurrency)}} $\
$\ $\\

\subsection{Micro-benchmark and Basic Prediction}
\label{sec:predict}
In the case of the GPUs we examine in this paper when the input size is large enough, the bottleneck of reduction algorithm is device memory bandwidth. Hence we use a memory bandwidth micro-benchmark to proxy the performance of reduction. To make this micro-benchmark an accurate representation, we add two add instructions to imitate the real computation in the reduction operation. Figure~\ref{fig:bandcode} shows the main instruction in the micro-benchmark.

\begin{figure}[t]
\centering
\begin{lstlisting}[language=C,gobble=2]
	while(i<n){sum+=g_idata[i];i+=groupsize;}
\end{lstlisting}
\caption{\label{fig:bandcode}Code example of the main instruction in the memory bandwidth micro-benchmark for proxying the reduction operation}
\end{figure}

Our objective is to identify when to use a single thread, a single warp barrier, and until when would it be more efficient to use a multi-GPU barrier. Instead of enumerating every possible case, we only consider two configurations here (and it can be extended to other cases):

\begin{itemize}
    \item To use a single thread or single warp barrier
    \item To use a single block with $1024$ threads or with $32$ threads
\end{itemize}


\begin{table} [t]
\caption{\label{tbl:bdwith} Predicting the switching point between two configurations}
\centering
\begin{tabular}{l|l|ll|ll}
\multicolumn{2}{l|}{\textbf{scenery}}&\multicolumn{2}{l|}{\textbf{sync ltc*}}&\multicolumn{2}{l}{\textbf{switch point}}\\
\multicolumn{2}{l|}{\textbf{}}&\multicolumn{2}{l|}{\textbf{cycle}}&\multicolumn{2}{l}{\textbf{B}}\\\hline
\multicolumn{2}{l|}{}      &V100&P100&V100&P100\\\hline
1           & 1 warp $N_l$   & \hfil 110 & \hfil 155 	&\hfil 70 &\hfil 70\\
            &1 warp $N_m$ &\hfil-        &\hfil- 		&\hfil 76 &\hfil 75\\\hline
2            &1024 thrd $N_l$&\hfil 420  &\hfil 2135 	&\hfil 9076 &\hfil 32681\\
            &1024 thrd $N_m$&\hfil -             &\hfil -           	&\hfil 8501 &\hfil 29737\\\hline
\end{tabular}
$_{\mathit{*:\ 5\ times\ synchronization}} $\
\end{table}

Normally in the two configurations we mentioned, the data is usually kept in shared memory or cache, so we only measure shared memory for the following part. Table \ref{tbl:reduce_concurency} shows the results of bandwidth (throughput), latency and concurrency.

Take the double type as an example ($8$ Bytes). In this case, in both configurations, the input size exceeds the concurrency of both "basic" and "more" settings, hence we only need to take $N_l$ in  Equation~\ref{eqt:predictsize_l} into consideration. Table~\ref{tbl:bdwith} shows the results.

Table \ref{tbl:bdwith} shows that: first, it is better to compute $32$ data points with a warp; second, there would be no benefit to compute $1024$ data points with $1024$ threads per block. Our further experiments show that those predictions are correct.

In addition, another potential overhead caused by synchronization would be that the synchronization would possibly clear the instruction pipeline. Threads might need additional time to saturate the pipeline. So the real switching point would likely be larger than this. 


\subsection{Warp Level Reduction}\label{sec:warpreduce}
In this subsection, we compare different warp level synchronization methods in the reduction kernel by observing their behaviour in the current generations of GPUs. Figure~\ref{fig:warp_reduction_code} shows our sample code, and Table~\ref{tbl:warp_reduction} shows the result.

\begin{figure}[t]
\centering
\begin{lstlisting}[language=C,gobble=2]
	//assume the data resides in shared memory
	for(step = 16; step >=1; step/=2){
		//or use the shuffle operation here
		if(tid+step<32)sm[tid]+=sm[tid+step];
		synchronize();
	}
\end{lstlisting}
\caption{\label{fig:warp_reduction_code}Code example of warp level reduction with synchronization}
\end{figure}

As shown in~Table~\ref{tbl:warp_reduction}, when using the \textit{volatile} qualifier for the input data, the performance of warp level synchronization is no worse than in the case without the volatile qualifier (shown as "tile" in the table). Accordingly, the warp level synchronization does not have much overhead other than to ensure memory consistency. We can conclude that warp level synchronization is no more than a memory fence in the current version of CUDA. We also observe that the results for using the shuffle operation with the tile group have the lowest latency.

\begin{table}[t]
\caption{\label{tbl:warp_reduction} Latency (cycles) to Compute Sum of $32$ values (double precision)}
\centering
\begin{tabular}{l|l|l|l|l|l|l|l}
& serial  & \hfil nosync & volatile  & tile &coa &tile& coa  \\ 
&  &\hfil * & \hfil \& tile & & &shuffle& shuffle  \\ \hline
\hfil V100&\hfil 299     & \hfil 89      & \hfil 237      & \hfil 237 & \hfil 237 & \hfil 164& \hfil 1261       \\ \hline
\hfil P100&\hfil 383     & \hfil 112     & \hfil 282      & \hfil 281 & \hfil 251 & \hfil 212& \hfil 1423  \\ \hline
\end{tabular}
$_{*\mathit{result\ of \ no\ synchronization\ version\ is\ incorrect}} $\
\end{table}

\subsection{Single GPU Reduction}
\label{sec:single_red}
In this Subsection, we directly apply the knowledge in Section~\ref{sec:predict} in implementing device-wide reduction. Figure~\ref{fig:gridreduce} shows the code of reduction with explicit synchronization and Figure~\ref{fig:kernelreduce} shows the code of reduction with implicit synchronization for a single GPU.

\begin{figure}[t]
\centering
\begin{lstlisting}[language=C,gobble=2]
	__device__ REAL summing(...){...
		uint i = threadid + blockid * blockdim;
		sum=0;
		while(i<n){
			sum+=g_idata[i]; 
			i+=blockdim*griddim;
		}
		return sum;
	}
	__device__ REAL block_reduce(...){...
		i = threadid;
		sum=0;
		while(i<n){sum+=td[i]; i+=blockdim;}
		//n is the pre-computed switch point
		td[threadid]=sum;
		sum=0;
		block.sync();
		if(warpid==0)
		{
			i = threadid;
			while(i<blockDim){sum+=td[i];i+=32;}
			sum = shuffle_reduce_warp(sum);
		}
		return sum;
	}
\end{lstlisting}
\caption{\label{fig:gridreducefunction} Basic function of device wide reduction}
\end{figure}

\begin{figure}[t]
\centering
\begin{lstlisting}[language=C,gobble=2]
	 //works in both single and multi GPU
	__global__ void ExplicitGPU(...){...
		while(step.notfinish()){
			//directly store data in the target GPU
			dest[step][threadid] 
			    = summing(src[step][threadid], ...); 
			grid.sync();//explicit synchronize;
		}
		if(gpu_id==0)
		{
			sum=block_reduce(src[0][0], ...);
			if(threadid==0)
				output[threadid]=sum;
		}
	}
\end{lstlisting}
\caption{\label{fig:gridreduce} Code example of reduction with explicit device synchronization}
\end{figure}

\begin{figure}[t]
\centering
\begin{lstlisting}[language=C,gobble=2]

	__global__ void Kernel1(...){...
		uint i = threadid + blockid * blockdim;
		sum=summing(...);
		output[i]=sum;
	...}
	__global__ void Kernel2(...){...
		sum=block_reduce(...);
		if(threadid==0)
			output[threadid]=sum;
	...}

	//following parts are CPU functions    
	void implicitSingleGPU(...){...
		Kernel1<<<...>>>(...);//implicit synchronization
		Kernel2<<<...>>>(...);
		...}

	void implicitMultiGPU(){...
	#pragma omp for num_threads(gpucount){...
		cudaDeviceSet(tid);
		Kernel1<<<...>>>(...);
		//gather data to one GPU that would do the remaining computation.
		while(step.notfinish()){
			cudaDeviceSynchronize();
			#pragma omp barrier;
			//transfer data from current GPU to another GPU
			transferdata(src[step][tid],dst[step][tid]);
		}
		cudaDeviceSynchronize();
		#pragma omp barrier;
		if(tid==0)Kernel2<<<...>>>(...);
        }
    ...}    
\end{lstlisting}
\caption{\label{fig:kernelreduce} Code example of reduction with implicit device synchronization}
\end{figure}

The widely used GPU C++ library CUB~\cite{nvidia2019cub} and CUDA SDK samples \cite{nvidia2019sample} include single GPU reduction implementations, we compare the performance of those implementations with our implementation.

\begin{figure}[t]
\centering
\begin{minipage}{0.48\textwidth}
\centering
\includegraphics[width=1\textwidth]{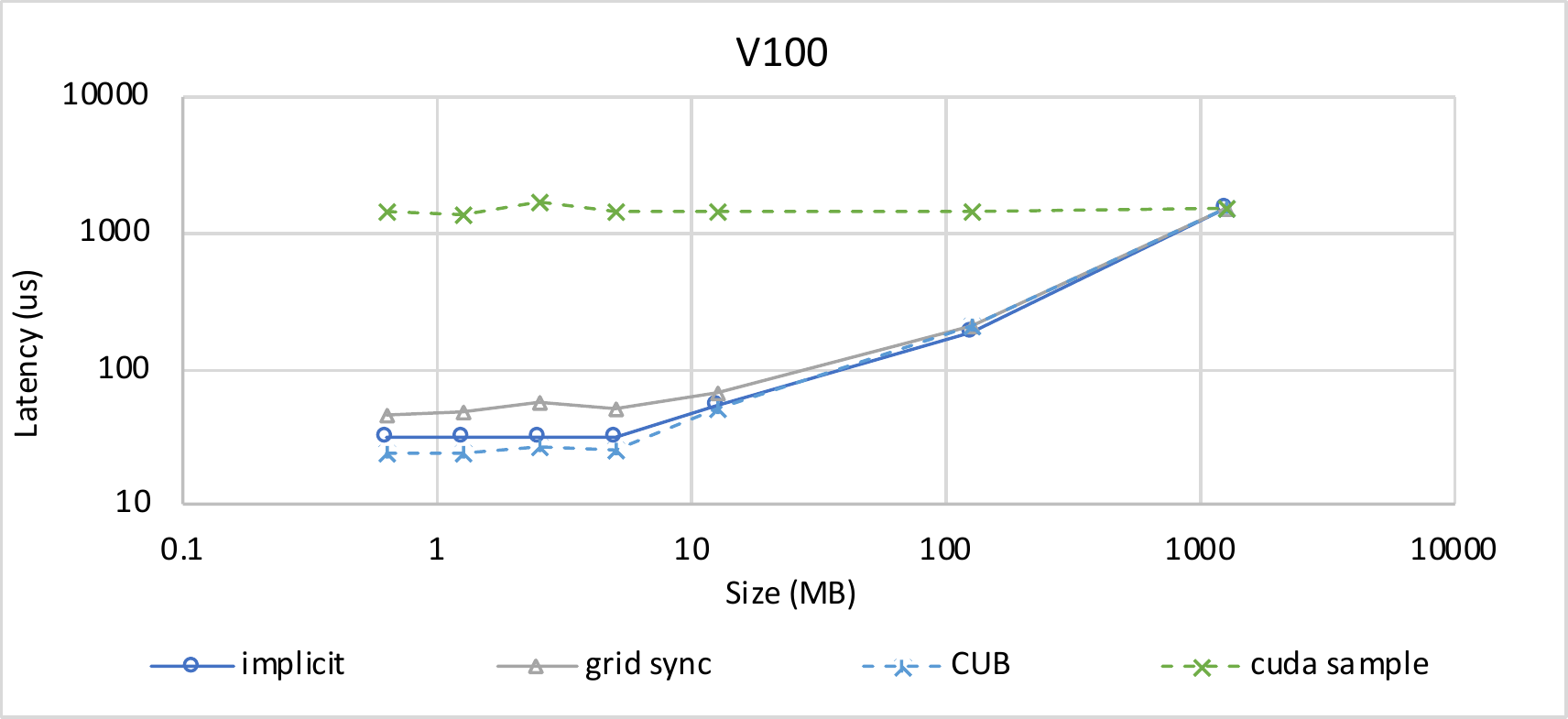}
\end{minipage}
\begin{minipage}{0.48\textwidth}
\centering
 \includegraphics[width=1\textwidth]{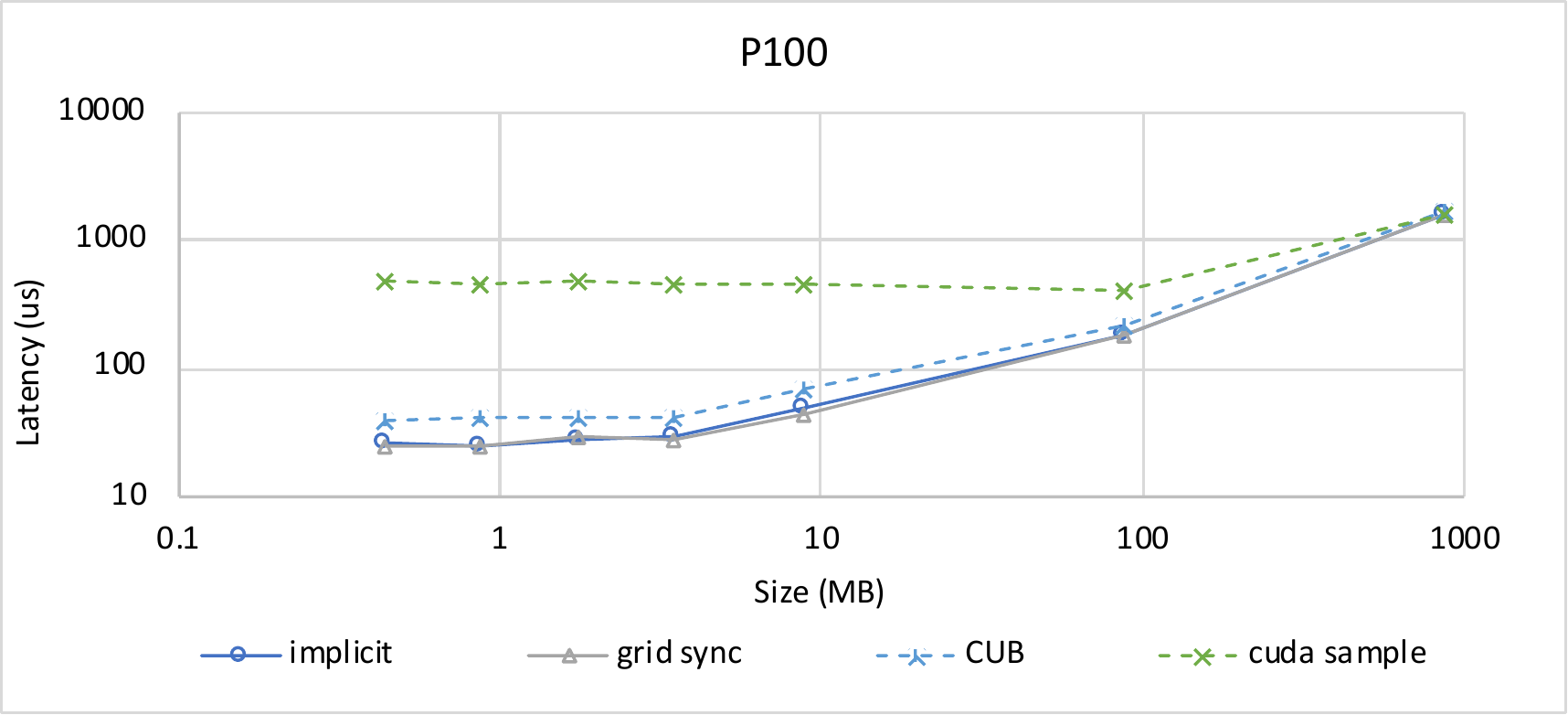}
\end{minipage}
\caption{Comparison of the performance of single reduction in V100 (up) and in P100 (down)}
\label{Fig:grid_reduce_rst}
\end{figure}
 
Figure~\ref{Fig:grid_reduce_rst} and Table~\ref{tbl:grid_reduction_bandwidth} show the results. Our implementation is comparable to state of art implementations on V100 and is noticeably better on P100. We can learn from Figure~\ref{Fig:grid_reduce_rst} that using a CPU-side barrier ("implicit" in the figure) always outperforms using grid synchronization ("grid sync" in the figure), though the performance difference is not so decisive. 

\begin{table}[t]
\caption{\label{tbl:grid_reduction_bandwidth} Bandwidth (GB/s) of different reduction methods}
\centering
\begin{tabular}{l|l|l|l|l|l}
&\hfil  implicit  & \hfil grid sync & \hfil CUB  & \hfil CUDA sample &\hfil  theory  \\ \hline
\hfil V100&\hfil   865.40  & \hfil 855.59      & \hfil 849.39      & \hfil 852.98 & \hfil  898.05   \\ \hline
\hfil P100&\hfil 592.40    & \hfil 590.85     & \hfil 543.96      & \hfil 590.65 & \hfil 732.16 \\ \hline
\end{tabular}
\end{table}

\subsection{Multi-GPU Reduction}
\label{sec:mul_red}
\begin{figure}[t]
\centering
\includegraphics[width=0.47\textwidth]{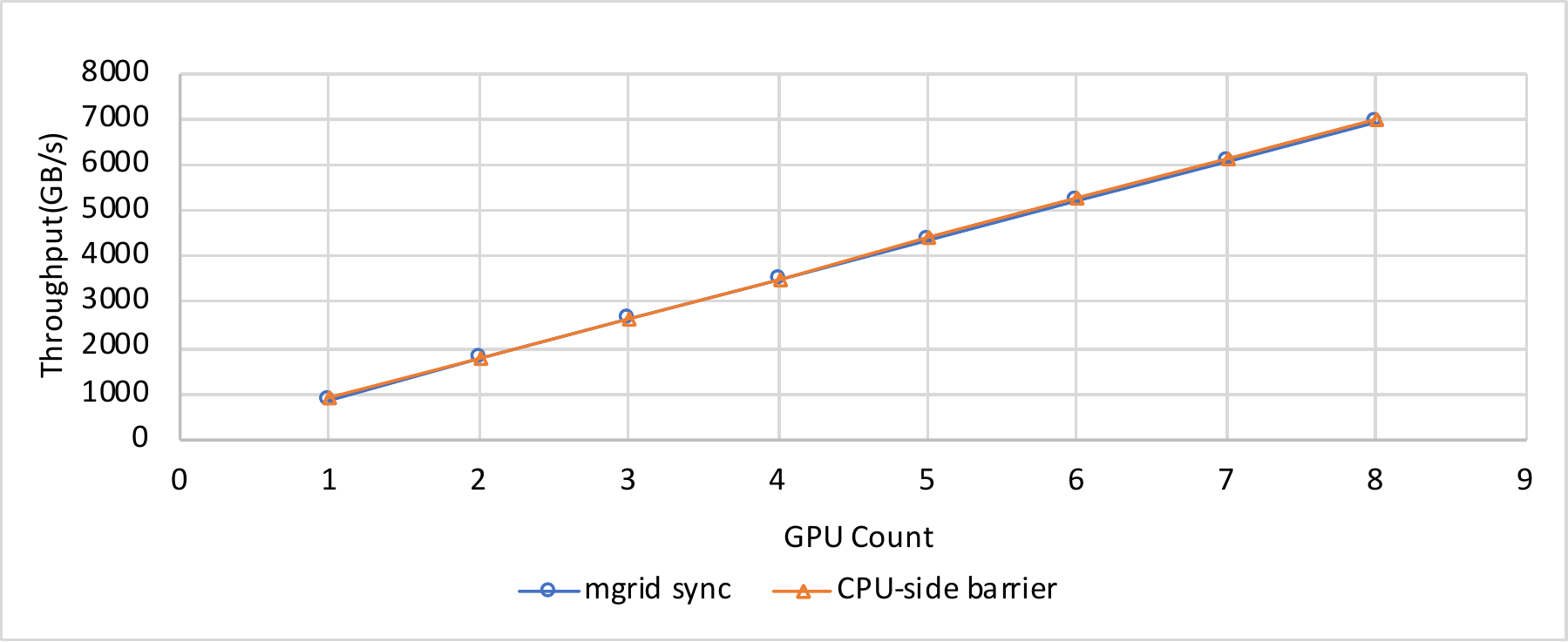}
\caption{The throughput of reduction on DGX-1}
\label{Fig:dgx_reduce}
\end{figure}

In this section, we use the code in Figure~\ref{fig:gridreduce} and implicitMultiGPU code in Figure~\ref{fig:kernelreduce}. Figure~\ref{Fig:dgx_reduce} shows the results. Though it is hard to notice, an implicit barrier is always slightly better than the multi-grid synchronization method. As section~\ref{sec:implicit} mentioned, the overhead in cooperative multi-launch might be the cause of this performance difference.


On the other hand, we want to emphasize here the benefit of programming. We can easily rewrite implicit barrier code (Figure~\ref{fig:kernelreduce}) into the explicit barrier one (Figure~\ref{fig:gridreduce}), i.e. a single persistent kernel is required in grid synchronization, and eliminate the complexity of managing several GPUs with CPU threads or processes. More importantly, the kernel function requires no knowledge of the hardware structure. 

\section{Considerations of Using CUDA Synchronization Instructions}
In this study, we identified several cases at which the synchronization instructions might not work as intended. In this section, we summarize some of those cases.
\subsection{Synchronization Inside a Warp}
\label{sec:warp_sc}
In this section, we examine synchronization at the warp level. To see if a barrier inside a warp is effective on all threads in the barrier, we run the code in Figure~\ref{fig:sc}. In the ideal case, the timers in all threads in the warp before the barrier are smaller than the timers after the sync in every thread. We test all the synchronization methods. Results show that P100 does not assure all threads inside a warp are blocked at the barrier (also the shuffle operation does not work correctly in this code either), which we believe explains why the latency of warp level synchronization in P100 is as fast as Table \ref{tbl:warprst} shows. On the other hand, in V100, we observed the anticipated behavior (likely due to the fact that in V100 each thread has its own program counter). Figure \ref{Fig:tile_timer} shows our observation when calling tile synchronization. We observed the same phenomenon when running all other synchronization instructions in both V100 and P100.  

\begin{figure}[t]
\centering
\begin{lstlisting}[language=C,gobble=2]
	if(tid==0){timer(start);sync;timer(end);}
	else if(tid==1){timer(start);sync;timer(end);}
	...
	else if(tid==30){timer(start);sync;timer(end);}
	else{timer(start);sync;timer(end);}
\end{lstlisting}
\caption{\label{fig:sc}Code example to verify synchronization inside a warp}
\end{figure}

\begin{figure}[t]
\centering
\begin{minipage}{0.24\textwidth}
\centering
\includegraphics[width=1\textwidth]{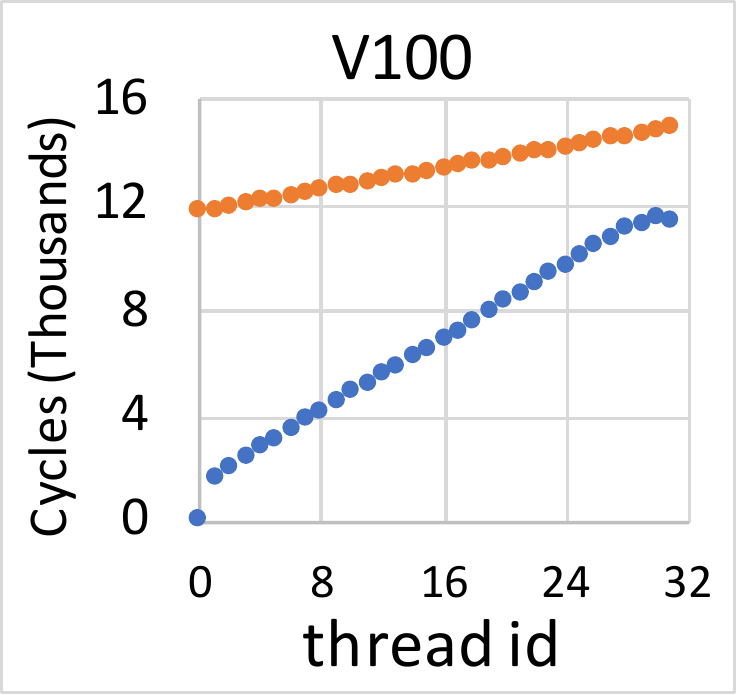}
\end{minipage}
\begin{minipage}{0.24\textwidth}
\centering
 \includegraphics[width=1\textwidth]{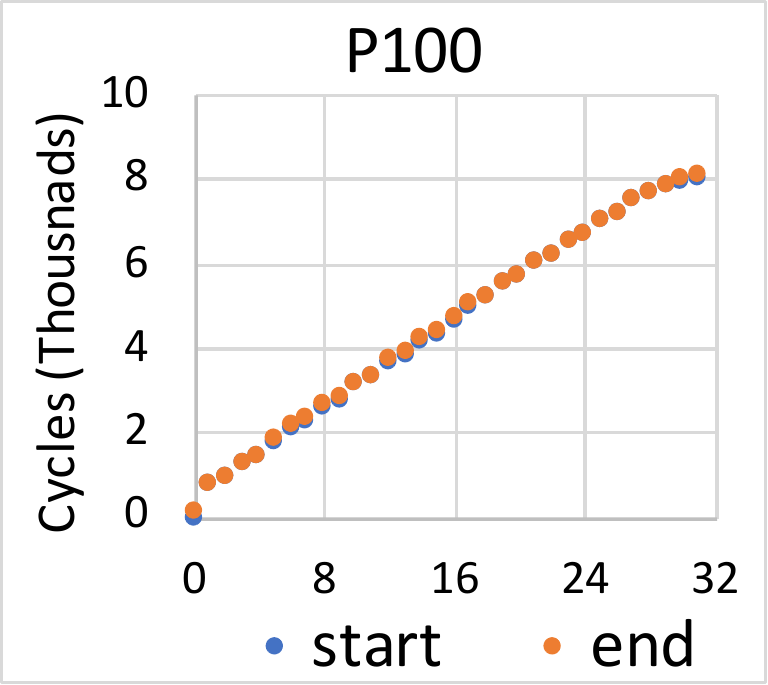}
\end{minipage}
\caption{Timer of threads inside a warp when calling tile synchronization in V100 (left), and in P100 (right) in code sample of Figure~\ref{fig:sc}}
\label{Fig:tile_timer}
\end{figure}
\subsection{Deadlocks in Synchronization of Parts of Thread Groups}
In this section, we examine the behaviour of synchronization with a subset of a thread group: would synchronizing a subset of a group cause a deadlock or not? We implement a test suite to see what happens when part of a thread group calls the synchronization function. We test through every granularity including threads, warps, blocks, and GPUs. As a result, we observed deadlocks when we synchronize parts of blocks in grid group, multi-grid group, and when we synchronize parts of GPUs in a multi-grid group. In summary, one should be careful, after initializing a grid group or a multi-grid group, since current CUDA does not support synchronizing sub-groups inside a grid group. 






\section{Benchmarking CUDA Synchronization Methods}

\label{sec:microbenchmark}

\subsection{Experiments Environment}
We use Pascal P100 and Volta V100 cards to conduct our experiments. 
We set the application frequency of both platforms to default. We use the latest stable driver. Table \ref{tbl:environment} shows the details of the environment.

\begin{table}[t]
\caption{\label{tbl:environment}Environment Information}
\centering
\begin{tabular}{l|l|l|l}
\hfil \textbf{Platform} & \hfil \textbf{Default Freq.} & \hfil \textbf{Driver }&\hfil \textbf{CUDA }\\\hline
\hfil P100 x 2& \hfil 1189MHz & \hfil 418.40.04& V10.0.130\\\hline
\hfil V100 x 8(DGX-1)& \hfil 1312MHz & \hfil 410.129& V10.0.130\\\hline
\end{tabular}
\end{table}

\subsection{Micro-benchmark for Implicit Barriers}
\label{sec:benchimplicit}

We use the terminologies in Section~\ref{sec:implicit}.
We do a warm-up kernel call before every measurement that we don't report the results for.

We found that directly using a null kernel would not give a correct result here. Because at this point the stream pipeline is not saturated enough: the overhead tested would be larger than usual. The kernel execution latency needs to be larger than a certain number. This value is around $5 us$ for a single GPU and around $250 us$ for $8$ GPUs in DGX-1. In order to control the kernel latency, we use the sleep instruction introduced in CUDA for Volta platform. We use kernel fusion to unveil the overhead hidden in kernel latency. The basic assumption here is that merging the work of multiple argument-less kernels into one single kernel does not introduce additional launch overhead, and then the time saved when using kernel fusion should be equal to the overhead of launching an additional kernel. From our previous observations, the sleep instruction has insignificant overhead and fits well into this assumption. In this situation, we can compute the overhead with Equation~\ref{eqt:lgovh}. 
    
Since we use the sleep instruction as a tool to analyze launch overhead, which is only available in Volta Platform in CUDA, we only conduct experiments on the V100 GPU for this experiment.

    \begin{equation}
    \begin{aligned}
    O &= \tfrac{Latency_{ij} - Latency_{ji}}{i-j}
    \end{aligned}
    \label{eqt:lgovh}
    \end{equation}
    $_{*\mathit{(O\ represents\ Overhead;\ In\ Latency_{ij}\ (the\ left\ one),}}$ \\
    $_{\mathit{i\ represents\ call\ launch\ function\ i\ times,}}$ \\ 
    $_{\mathit{j\ represents\ launch\  kernels\ with\ j\ wait\ unit)}}$  \ 
    $\ $\\

To the best of the authors' knowledge, Volkov et al.~\cite{volkov2008benchmarking} was the first one measured the overhead of implicit barrier, i.e. CUDA kernel launch overhead. Xiao et al.~\cite{xiao2010inter} additionally build a model for implicit and explicit barriers. They both neglect the fact that the launch overhead is far smaller when kernel execution latency is long enough. When using null kernels, we tested a launch overhead of around $3 us$ for traditional launch, which is the same as the best case reported by Volkov et al.~\cite{volkov2008benchmarking}. 

\subsection{Micro-benchmark for Intra-SM Instructions}
\label{sec:benchintrasm}
\label{sec:gpubase}
We directly use Wong's \cite{wong2010demystifying} method for instruction micro-benchmarking. Wong's method relies on the GPU clock. The basic methodology is to build a chain of dependent operations to repeat a single instruction enough times to saturate the instruction pipeline. By using the clock register to record the begin and end timestamps of the series of operations, it is possible to average the repetitions to infer the latency of that instruction. Figure~\ref{Fig:gputest} shows an example sample code to measure the latency of an add instruction.  

\begin{figure}[t]
\centering
\begin{lstlisting}[language=C,gobble=2]
	__global__ void kernel1(){
		start=clock();
		repeat256(p=p+q;q=p+q); //repeat=512
		end=clock();
		return q;
		}
\end{lstlisting}
\begin{lstlisting}[language=C,gobble=2]
	__global__ void kernel2(){
		start=clock();
		repeat512(p=p+q;q=p+q); //repeat=1024
		end=clock();
		return q;
		}
\end{lstlisting}

\begin{lstlisting}[language=C,gobble=2]
	cpuclock();
	kernel();
	syncdevice();
	cpuclock();
\end{lstlisting}
\caption{\label{Fig:gputest} Sample code to measure the latency of the add instruction in GPU}
\end{figure}

\subsection{Micro-benchmark for Inter SM Instructions}
\label{sec:benchintersm}
\label{sec:cpubase}
Jia's work \cite{jia2018dissecting} can work correctly only inside a single thread, Wong's work \cite{wong2010demystifying} can work correctly only in a single SM. Yet current synchronization instructions might involve cooperation across different threads, different SMs, and even different GPUs. As we move to grid level synchronization and beyond, we need a new method.

In order to test the performance of synchronization beyond a single SM, a global clock is necessary. In CUDA's execution model, a CPU thread launches a kernel and it can call the DeviceSynchronize() function to block the CPU thread until the GPU kernel finishes execution. So it is possible to use the clock in that CPU thread as a global clock to test GPU instructions. Yet we need to fix two issues before we can use the CPU clock:

\begin{itemize}
\item We need to eliminate any latency not related to the target instruction
\item Account for the relative inaccuracy in the CPU clock measurement, in comparison to the GPU's clock measurement.
\end{itemize}

In order to solve those issues, we need to additionally introduce two assumptions:
\begin{itemize}
\item The measurement of the latency of every instruction becomes more accurate when the pipeline is saturated
\item Additional instructions in a kernel do not increase the launch overhead of kernel launch
\end{itemize}

\begin{equation}
    T_{instruction} = \tfrac{{L_{k_1} - {L_{k_2}}}} {r_{1}-r_{2}}
    \label{eqt:ltc}
\end{equation}
\begin{equation}
	\begin{split}
    \sigma_{\tfrac{k_1-k_2}{r_1-r_2}}& =\sqrt{\tfrac{\sum_{n=1}^{N}{(\tfrac{L_{k_1}-L_{k_2}}{r_1-r_2})}^2-\sum_{n=1}^{N}{ \overline{(\tfrac{L_{k_1}-L_{k_2}}{r_1-r_2})}}^2}{N-1}}\\
   & =\tfrac{1}{r_1-r_2}\sqrt{\tfrac{\sum{L_{k_1}^2- \overline{L_{k_1}}^2}}{N-1}+\tfrac{\sum{L_{k_2}^2-\overline{L_{k_2}}^2}}{N-1}} \\
   & =\tfrac{1}{r_1-r_2}\sqrt{\sigma_{k_1}^2+\sigma_{k_2}^2}
   \end{split}
    \label{eqt:stdori}
\end{equation}
    $_{*\mathit{(L_{k_i}\ represents\ kernel\ total\ latency\ of\ kernel\ i;}}$ \\
    $_{\mathit{r_i\ represents\ repeat\ times\ in\ kernel\ i)}}$ \
    $\ $\\

Under those assumptions, if we increase the repetitions of instructions in the GPU kernel (in Figure~\ref{Fig:gputest}), the additional kernel latency is only related to the additional repeat times of instructions. In this manner, we are able to avoid unrelated latency that might come from kernel launch (to get more accurate measurements). Equation~\ref{eqt:ltc} shows how to measure the instruction latency with this method. (First issue solved)

Standard deviation can be used to represent the uncertainty in a single measurement~\cite{taylor1997introduction}. Equation~\ref{eqt:stdori} shows the standard deviation of the instruction tested, and its deduction (the measurement of kernel 1 and kernel 2 is independent to each other). And by deduction, if the difference in repeat times is large enough, the standard deviation of the instruction latency we seek to measure will be small. (Second issue solved)



In order to verify that the method we proposed in Section~\ref{sec:cpubase} matches our assumptions, we use both Wong's method and our method to test the single precision add instruction. Both results show that float-add costs $6$ cycles in P100 and $4$ cycles in V100. Those results match the result in~\cite{jia2018dissecting}. We can conclude that the inter SM micro-benchmark method we propose is a reliable measurement tool that approaches the accuracy of the GPU clock.




We additionally verify that the repeat times of a synchronization instruction itself would not influence the performance itself in block and grid level. Tile shuffle at warp level also works as we anticipated. Other warp level synchronization can be unstable: the latency of the synchronization instruction might increase suddenly when increasing repeat times. It could be the case that this warp synchronization relies on a software implementation. So when repeating an instruction too many times, instruction cache overflow can occur. We only record the fastest result for warp level synchronization instructions.

\section{Conclusion}
In this paper, we conduct a detailed study of different synchronization methods in Nvidia GPUs, ranging from warp to grid, and from single GPU to multi-GPU. 

We find that the performance of block synchronization is related to the number of warps involved, and the performance of grid level synchronization is mainly affected by the number of blocks involved. In addition, the performance of multi-grid level synchronization depends on the network structure connecting the GPUs, and the number of active blocks and warps. 

CPU-side implicit barriers generally perform better than grid level and multi-grid level synchronization. But if the program size is large enough, the performance difference would not be so severe, with the added benefit that multi-grid synchronization simplifies multi-GPU programming.

We use the reduction operator as an example to use the knowledge we gain from micro-benchmark. We build a performance model to predict where would be the point that using fewer threads is more profitable. Additionally, using code samples, we show a possible simple way to do multi-GPU programming without much performance degradation. Moreover, with more multi-grid barriers in a kernel, the launch overhead in multi-device kernel launch would become more insignificant. Table~\ref{tbl:tips} summarizes the knowledge we gained from this study.  


\begin{table}[t]
\caption{\label{tbl:tips} Summary of Observations}
\centering
\begin{tabular}{p{0.12\textwidth}|p{0.3\textwidth}}\hline
\hfil \textbf{Warp Level Sync}&  Does not work on Pascal; \\ &Shuffle performs better in real code.\\\hline
\hfil \textbf{Block Sync }    & The number of active warps per SM affects performance\\\hline
\hfil \textbf{Grid Sync }     & The number of blocks per SM mainly affects performance;\\&Generally, the performance is acceptable if $block/SM<=2$;\\& Currently, only parts of blocks inside a grid calling grid level synchronization would cause deadlock.\\\hline
\hfil \textbf{Multi-Grid Sync} &  Both the number of blocks per SM and active warps per SM affect performance; \\& If $thread/SM <= 1024$ and $ block/SM <=  8$ the performance is relatively acceptable;\\& Currently, only parts of grids inside a grid calling grid level synchronization would cause deadlock.\\\hline
\hfil \textbf{Implicit Sync \& CPU Based Sync} &  Generally, their performance is slightly better than explicit synchronization when in single GPU or when the GPU count is large, or when there is no much synchronization steps; \\& The issue for CPU Based Sync is programmability, especially in the situation of multi-GPUs.\\\hline
\end{tabular}
\end{table}

\section{ACKNOWLEDGMENTS}
This work was partially supported by JST-CREST under Grant Number JPMJCR19F5.

\bibliographystyle{./bibliography/IEEEtran}
\bibliography{./bibliography/IEEEabrv,./bibliography/IEEEexample}

\end{document}